\documentclass[twocolumn,pre,aps,showpacs]{revtex4}
\usepackage{mathrsfs}
\usepackage{graphicx}
\begin{document}
\title {Exact temporal evolution of the two-species Bose-Einstein condensates}
\author{Cong Zhang}
\author{Zhi-Hai Zhang}
\author{Shi-Jie Yang\footnote{Corresponding author: yangshijie@tsinghua.org.cn}}
\affiliation{Department of Physics, Beijing Normal University, Beijing 100875, China}
\begin{abstract}
We construct exact stationary solutions to the one-dimensional coupled Gross-Pitaevskii equations for the two-species Bose-Einstein condensates with equal intraspecies and interspecies interaction constants. Three types of complex solutions as well as their soliton limits are derived. By making use of the $SU(2)$ unitary symmetry, we further obtain analytical time-evolving solutions. These solutions exhibit spatiotemporal periodicity.
\end{abstract}
\pacs{03.75.Mn, 03.75.Kk, 03.75.Lm}
\maketitle

\section{Introduction}
Bose-Einstein condensates (BECs) in weakly interacting atomic gases have offered a practical means of
studying nonlinear behavior by using the matter waves. Among the macroscopic excitations, solitons and vortices are the most common and are extensively studied both theoretically and experimentally. The repulsive BEC has received most experimental attentions while the attractive BEC is believed to only be stable in one-dimensional (1D) systems\cite{Ruprecht,Kivshar1}. The Gross-Pitaevskii equation (GPE), which describes the mean field dynamics of a dilute BEC, can be reduced to a quasi-1D nonlinear Schr\"{o}dinger equation (NLSE), with a limit that the transverse dimensions of the condensate are on the order of its healing length and the longitudinal dimension is much longer than its transverse ones.

The observation of solitons in BECs brings great progresses in the study of the exactly integrable nonlinear systems which have a variety of applications in fiber optics as well as other
fields\cite{Carr3,Kivshar2,Sulem,Carr2}. Dark solitons are expected
only to exist in repulsive interactions ($s$-wave scattering length $a>0$), whereas bright
solitons exist in attractive interactions ($a<0$). The 1D NLSE is exactly solvable by the inverse scattering transform\cite{Zakharov-1,Zakharov-2}, and has a countably infinite number of conserved
quantities\cite{Miura-1,Miura-2}. For the single-component BEC, the stationary solutions have been deduced
analytically under box or periodic boundary conditions\cite{Carr2,Carr1}.

A very relevant generalization of this class of physical systems is the coupled multiple-species BECs. Various soliton complexes including bound dark-dark\cite{Ohberg}, dark-bright\cite{Busch}, dark-antidark, dark-gray, bright-antidark, and bright-gray\cite{Kevrekidis} were predicted. In Ref.[\onlinecite{Deconinck}], the authors constructed solutions of the coupled NLSEs with additional linear Rabi coupling to the problem where this linear coupling is absent. In this paper we seek exact time-evolving solutions to the 1D coupled GPEs which describe the two-species BECs. For equal intraspecies and interspecies coupling constants, the equations of motion are an integrable Manakov system which is applicable to the BECs consisting of different hyperfine states of $^{87}$Rb. In this case, the Hamiltonian has the $SU(2)$ symmetry. The normalized GPEs for the mean-field order parameter $\Psi=(\psi_1,\psi_2)^T$ in the uniform external potential read,
\begin{equation}
i\partial_t\Psi(x,t)=-\frac{1}{2}\partial_x^2\Psi(x,t)+\gamma |\Psi(x,t)|^2\Psi(x,t),\label{GPE}
\end{equation}
where $\gamma$ is the intraspecies and interspecies coupling constant.

The stationary equations are obtained by substituting $\psi_i(x,t)=\psi_i(x)\exp(-i\mu_i t)$ ($i=1,2$) as
\begin{equation}
  \mu_i\psi_i(x)=-\frac{1}{2}\partial_x^2\psi_i(x) +\gamma (\sum_{j=1,2}|\psi_{j}(x)|^2) \psi_i(x).\label{stationary}
\end{equation}
It should be noted that the solutions to Eqs.(\ref{stationary}) are not really "stationary" because the time-dependent phase factor in each component is different as $\mu_1\neq\mu_2$. So there exists a Lamor procession around the $z$-axis in the pseudo-spin space. However, the density distribution of each component is time-invariant. Hence we briefly call them the stationary states. The periodic boundary conditions
\begin{equation}
\psi_i(1)=\psi_i(0),
\hspace{4mm}\psi_i^\prime(1)=\psi_i^\prime(0),\label{boundary}
\end{equation}
are applied and the wavefunctions are normalized to
\begin{equation}
\int (|\psi_1(x)|^2+|\psi_2(x)|^2)dx=1.
\end{equation}

We first present three types of complex-real solutions to the stationary equations (\ref{stationary}) and examine the corresponding soliton limits. By applying a unitary transformation to Eq.(\ref{GPE}), we can obtain exact time-evolving solutions.

\section{type A}
We first consider the following form of stationary solutions to Eq.(\ref{stationary}) by the combination of function $\textrm{sn}(kx,m)$ and $\textrm{cn}(kx,m)$ as
\begin{eqnarray}
\left\{\begin{array}{c}
\psi_1(x)=f(x)e^{i\theta(x)}\\
\psi_2(x)=D \textrm{sn}(kx,m),\label{complexcnsn}
      \end{array}\right.
\end{eqnarray}
where $f(x)=\sqrt{A+B\textrm{cn}^2(kx,m)}$ with $A$, $B$, $D$ the real constants. $\textrm{sn}$ and $\textrm{cn}$ are the Jacobian elliptical functions with modulus $0<m<1$. The period\cite{Bowman,Abramowitz} $k=4jK(m)$ with $K(m)$ is the complete elliptic integral of the first kind. In the following we always choose the number of periods $j=2$. By substituting the solution (\ref{complexcnsn}) into the stationary Eqs. (\ref{stationary}) and making use of the identity between the Jacobian elliptical functions $\textrm{sn}^2+\textrm{cn}^2=1$, we obtain the following decoupled equations\cite{Zhang,Liu}
\begin{eqnarray}
\left\{\begin{array}{c}
  \tilde\mu_1\psi_1(x)=-\frac{1}{2}\partial_x^2\psi_1(x) +\tilde\gamma_1 |\psi_1(x)|^2 \psi_1(x) \\
  \tilde\mu_2\psi_2(x)=-\frac{1}{2}\partial_x^2\psi_2(x) +\tilde\gamma_2 |\psi_2(x)|^2\psi_2(x),\label{decouple}
\end{array}\right.
\end{eqnarray}
where $\tilde\mu_{1,2}$ are the effective chemical potentials and $\tilde\gamma_{1,2}$ the effective intraspecies coupling constants in each species, respectively. One has
\begin{eqnarray}
\left\{\begin{array}{c}
\tilde\mu_1=\mu_1 -\gamma (A+B)\frac{D^2}{B}\\
\tilde\mu_2=\mu_2 -\gamma(A+B),\\
      \end{array}\right.
\end{eqnarray}
and
\begin{eqnarray}
\left\{\begin{array}{c}
\tilde\gamma_1=\gamma (1-\frac{D^2}{B})\\
\tilde\gamma_2=\gamma (1-\frac{B}{D^2}).
      \end{array}\right.
\end{eqnarray}

The decoupled Eqs.(\ref{decouple}) can be self-consistently solved to yield
\begin{eqnarray}
\left\{\begin{array}{c}
\mu_1=\frac{1}{2} k^2(1-2m^2)-\frac{3}{2} \widetilde{\gamma_1} A +\gamma (A+B)\frac{D^2}{B}\\
\mu_2=\frac{1}{2} k^2(1+m^2)+\gamma(A+B),\\
      \end{array}\right.
\end{eqnarray}
and
\begin{equation}
D^2=\frac{m^2k^2}{\gamma}+B.
\end{equation}

There are two cases according to the value of $B>0$ and $B<0$. For $B>0$, we have the effective couplings $\tilde\gamma_1<0$ and $\tilde\gamma_2>0$, independent of the sign of the real nonlinear interaction $\gamma$. For $B<0$, the nonlinear interaction $\gamma$ must be repulsive and the effective couplings of both species also should be repulsive ($\tilde\gamma_1>0,\tilde\gamma_2>0$).

\begin{figure}[htbp]
\begin{center}
\includegraphics*[width=8cm]{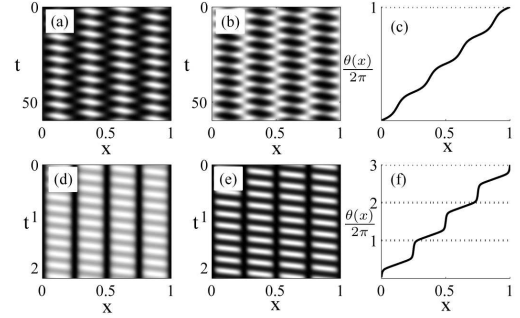}
\caption{Left two columns: temporal evolution of the density profiles for the state (\ref{psiA}). For the non-soliton cases (upper row), the parameters are $m=0.1$, $\gamma=100$,
$\mu_1=41170.48$, $\mu_2=41171.53$. For the soliton cases, $m=0.999$, $\gamma=0.01$, $\mu_1=1322.01$, and
$\mu_2=1293.58$. Right column: the phase profiles of the stationary state (\ref{complexcnsn})}.
\end{center}
\end{figure}

The phase $\theta(x)$ is determined by the imaginary part of the decoupled Eqs.(\ref{decouple}) which is given by
\begin{equation}
\theta(x)=\int_0^x\frac{\alpha}{f^2(\xi)}d\xi,\label{phase}
\end{equation}
where $\alpha=\pm(2\tilde{\mu}_1A^2-2\tilde{\gamma}_1A^3+k^2(1-m^2)AB)^{\frac{1}{2}}$
is an integral constant.

The periodic boundary conditions (\ref{boundary}) require
\begin{equation}
\theta(1)-\theta(0)=2j\pi\times n,
\end{equation}
where $n$ is an integer. The periodic condition for the phase $\theta(x)$ can be fulfilled by properly
adjusting the modulus $m$ of the Jacobian elliptic functions. Figure 1 (c) and (d) show the distribution of the phase for $m=0.1$ and $m=0.999$, respectively. The Fig.1(d) and (f) are the phase profiles $\theta(x)$ of stationary state $\Psi$. The monotonous gradient in the phase profile indicates that a supercurrent is carried by $\psi_1$.

We can further obtain the exact time-evolving solutions to Heisenberg equations of motion (\ref{GPE}) by making use of the $SU(2)$ symmetry of the Hamiltonian. We consider an arbitrary pseudo-spin rotational transformation $\Psi^\prime(x,t)=U(\phi)\Psi(x,t)$ with\cite{Deconinck}
\begin{equation}
U(\phi)=
\left(
\begin{array}{ccc}
 \cos\phi & \sin\phi \\
 -\sin\phi & \cos\phi\label{transfor}
\end{array}
\right),
\end{equation}
where the time factor $e^{-i\mu_i t}$ is attached to each component. Consequently, $\Psi^\prime(x,t)$ is also a solution to Eq.(\ref{GPE}) which is explicitly written as
\begin{widetext}
\begin{eqnarray}
\left\{\begin{array}{c}
\psi^\prime_1(x,t)=\exp(-i\mu_2 t)[\sqrt{A+B\textrm{cn}^2(kx,m)}e^{i\theta(x)} \cos\phi \exp(-i\mu t)+D\textrm{sn}(kx,m) \sin\phi],\\
\psi^\prime_2(x,t)=\exp(-i\mu_2 t)[-\sqrt{A+B\textrm{cn}^2(kx,m)}e^{i\theta(x)} \sin\phi \exp(-i\mu t)+D\textrm{sn}(kx,m) \cos\phi],
      \end{array}\right.\label{psiA}
\end{eqnarray}
\end{widetext}
where $\mu=\mu_1-\mu_2$ is the temporal evolving frequency. Obviously, the density distribution possesses the spatiotemporal periodicity.

Figure 1 shows the temporal density evolution of the state $\Psi^\prime(x,t)$ (\ref{psiA}). Hereafter we always choose the rotational angle $\phi=\pi/8$. Fig.1(a-c) are for $m=0.1$ and the corresponding parameters are $\gamma=100$, $A=101.38$, $B=309.5$, $D^2=309.6$, and $n=1$. Fig.1(d-f) are for $m=0.999$ which correspond to the soliton limit. The relevant parameters are $\gamma=0.01$, $A=3.3\times 10^3$, $B=-3.2\times 10^3$, $D^2=1.26\times 10^5$, and $n=3$. It displays a dark-gray solitonic complex.

\section{type B}
By the same way, we construct the following form of complex solutions
\begin{eqnarray}
\left\{\begin{array}{c}
\psi_1(x)=f(x)e^{i\theta(x)}\\
\psi_2(x)=D \textrm{cn}(kx,m),\label{complexcncn}
      \end{array}\right.
\end{eqnarray}
where $f(x)=\sqrt{A+B\textrm{cn}^2(kx,m)}$ with $A$, $B$, $D$ the real constants. The effective chemical potentials $\tilde\mu_{1,2}$ and effective coupling constant $\tilde\gamma_{1,2}$ are obtained as
\begin{eqnarray}
\left\{\begin{array}{c}
\tilde\mu_1=\mu_1 +\gamma A\frac{D^2}{B}\\
\tilde\mu_2=\mu_2- \gamma A,\\
      \end{array}\right.
\end{eqnarray}
and
\begin{eqnarray}
\left\{\begin{array}{c}
\tilde\gamma_1=\gamma (1+\frac{D^2}{B})\\
\tilde\gamma_2=\gamma (1+\frac{B}{D^2}).
      \end{array}\right.
\end{eqnarray}
The relevant parameters are
\begin{eqnarray}
\left\{\begin{array}{c}
\mu_1=\frac{1}{2} k^2(1-2m^2)-\frac{3}{2} \widetilde{\gamma_1} A -\gamma A\frac{D^2}{B}\\
\mu_2=\frac{1}{2} k^2(1-2m^2)+\gamma A,\\
      \end{array}\right.
\end{eqnarray}
and
\begin{equation}
D^2=-\frac{m^2k^2}{\gamma}-B.
\end{equation}
The phase $\theta(x)$ is also expressed by Eq.(\ref{phase}). Figure 2(c) and (f) show the phase profiles of the state (\ref{complexcncn}) for $m=0.1$ and $m=0.999$, respectively. It indicates a super current in the condensate.

There are also two cases for $B>0$ and $B<0$. For $B>0$, we have $\gamma<0$ and the effective couplings of the two species should be attractive, $\tilde\gamma_1<0$ and $\tilde\gamma_2<0$. For $B<0$, we note that the effective coupling in the first species is repulsive ($\tilde\gamma_1>0$) while the effective coupling in the second species is attractive ($\tilde\gamma_2<0$), independent of the sign of the real nonlinear interaction $\gamma$.

\begin{figure}[htbp]
\begin{center}
\includegraphics*[width=8cm]{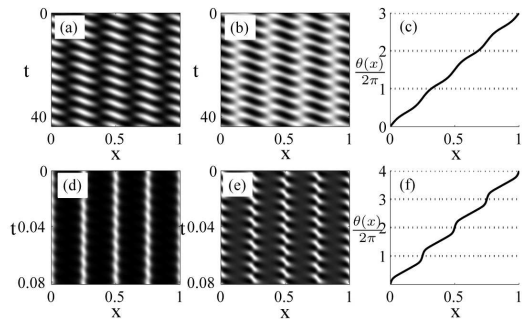}
\caption{Left two columns: temporal evolution of the density profiles for another state (\ref{psiB}). Here are the given parameters: (a,b,c) the non-soliton state shown in upper row corresponds to $m=0.1$, $\gamma=0.1$, $\mu_1=130.90$, $\mu_2=129.51$; (d,e,f) the soliton cases corresponds to $m=0.999$, $\gamma=-0.1$, $\mu_1=22.3771$ and $\mu_2=-747.29$. Right column: the phase profiles of the stationary state (\ref{complexcncn}).}
\end{center}
\end{figure}

After pseudo-spin rotational transformation, we again work out the time-evolving solution to the Heisenberg equations of motion (\ref{GPE}) for the two species as
\begin{widetext}
\begin{eqnarray}
\left\{\begin{array}{c}
\psi^\prime_1(x,t)=\exp(-i\mu_2 t)[\sqrt{A+B\textrm{cn}^2(kx,m)}e^{i\theta(x)} \cos\phi \exp(-i\mu t)+D\textrm{cn}(kx,m) \sin\phi],\\
\psi^\prime_2(x,t)=\exp(-i\mu_2 t)[-\sqrt{A+B\textrm{cn}^2(kx,m)}e^{i\theta(x)} \sin\phi \exp(-i\mu t)+D\textrm{cn}(kx,m) \cos\phi],
      \end{array}\right.\label{psiB}
\end{eqnarray}
\end{widetext}
where $\mu=\mu_1-\mu_2$.

Figure 2 shows the temporal density evolution of the state $\Psi^\prime$ (\ref{psiB}). Fig. 2(a-c) correspond to non-soliton cases $m=0.1$ with the attractive coupling $\gamma=0.1$, $A=517.5$, $B=-287.26$, $D^2=271.3911$ and $n=3$, while Fig. 2(d-f) are for the soliton limits $m=0.999$ with repulsive coupling $\gamma=-0.1$, $A=1031.5$, $B=-865$, $D^2=1.3774\times 10^4$ and $n=4$. Under this circumstances, we obtain a bright-bright solitonic complex in the condensate.

\section{type C}
We can construct the third type of complex solutions
\begin{eqnarray}
\left\{\begin{array}{c}
\psi_1(x)=f(x)e^{i\theta(x)}\\
\psi_2(x)=D \textrm{dn}(kx,m).\label{complexcndn}
      \end{array}\right.
\end{eqnarray}
The effective chemical potentials and effective coupling constants are respectively
\begin{eqnarray}
\left\{\begin{array}{c}
\tilde\mu_1=\mu_1-\gamma D^2(1+m^2)-\gamma A\frac{m^2 D^2}{B}\\
\tilde\mu_2=\mu_2-\gamma A-\gamma B\frac{m^2-1}{m^2},
      \end{array}\right.
\end{eqnarray}
and
\begin{eqnarray}
\left\{\begin{array}{c}
\tilde\gamma_1=\gamma (1+\frac{m^2 D^2}{B})\\
\tilde\gamma_2=\gamma (1+\frac{B}{m^2 D^2}).
      \end{array}\right.
\end{eqnarray}

The decoupled equations (\ref{decouple}) are self-consistently solved to yield
\begin{eqnarray}
\left\{\begin{array}{c}
\mu_1=\frac{1}{2} k^2(1-2m^2)-\frac{3}{2} \widetilde{\gamma_1} A+\gamma D^2(1+m^2)+\gamma A\frac{m^2 D^2}{B}\\
\mu_2=\frac{1}{2} k^2(m^2-2)+\gamma A+\gamma B\frac{m^2-1}{m^2},\\
      \end{array}\right.
\end{eqnarray}
and
\begin{equation}
B+m^2 D^2=-\frac{m^2 k^2}{\gamma}.
\end{equation}

The signs of the effective couplings in the two species are similar to those in the TYPE B solution.

\begin{figure}[htbp]
\begin{center}
\includegraphics*[width=8cm]{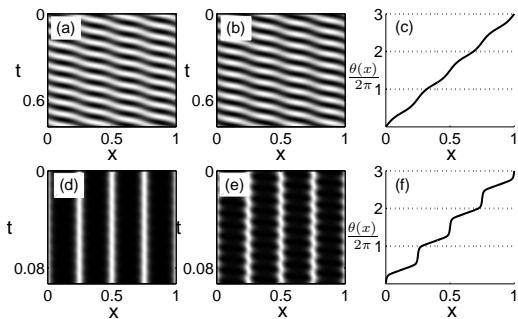}
\caption{Left two columns: temporal evolution of the density profiles for state (\ref{psiC}).
Parameters in the non-soliton state cases (upper row) of $m=0.1$ are $\gamma=-0.1$, $\mu_1=3033$, and $\mu_2=2950$. The parameters of the soliton cases of $m=0.999$ are $\gamma=-0.1$, $\mu_1=21.15$ and $\mu_2=-654$. Right column:
the phase profiles of the stationary state (\ref{complexcndn}).}
\end{center}
\end{figure}

Through the pseudo-spin rotational transformation, we obtain the time-evolving solutions to the Eq.(\ref{GPE}) as
\begin{widetext}
\begin{eqnarray}
\left\{\begin{array}{c}
\psi^\prime_1(x,t)=\exp(-i\mu_2 t)[\sqrt{A+B\textrm{cn}^2(kx,m)}e^{i\theta(x)} \cos\phi \exp(-i\mu t)+D\textrm{dn}(kx,m) \sin\phi],\\
\psi^\prime_2(x,t)=\exp(-i\mu_2 t)[-\sqrt{A+B\textrm{cn}^2(kx,m)}e^{i\theta(x)} \sin\phi \exp(-i\mu t)+D\textrm{dn}(kx,m) \cos\phi],
      \end{array}\right.\label{psiC}
\end{eqnarray}
\end{widetext}
with the temporal evolving frequency $\mu=\mu_1-\mu_2$.

In Fig. 3, a graphical temporal evolution of the density profiles of this solution is displayed for $\gamma=-0.1$. Fig.3(a-c) demonstrate case of $m=0.1$ with the corresponding parameters are calculated as $A=555.5$, $B=-308.36$, $D^2=2.925\times 10^4$ and $n=3$. Figure 3(d-f) display the soliton limit with $m=0.999$. We obtain the parameters $A=60.268$, $B=-57.92$, $D^2=1.2992\times 10^4$ and $n=3$, respectively. The condensate also forms a bright-bright solitonic complex. The phase profiles $\theta(x)$ of stationary state $\Psi$ is displayed in Fig.3(d) and (f).

\section{Summary}
In summary, we obtained three types of stationary solution to the pseudo-spin-1/2 BEC which possesses the $SU(2)$ symmetry. Exact time-evolving solutions are constructed through a pseudo-spin rotational transformation. The states exhibit periodicity in both space and time. In the limit of $m\rightarrow 1$, we obtained the solitonic complexes for the two species condensate.

This work is supported by funds from the Ministry of Science and Technology of China under Grant No.
2012CB821403.

\end{document}